\documentclass[twocolumn,prl]{revtex4}

\usepackage{graphicx}% Include figure files 
\usepackage{dcolumn}% Aligntable columns on decimal point 
\usepackage{bm}% bold math
\usepackage{times}

\begin{document}

%\preprint{Ge preprint} 
\title{Induced Ge Spin Polarization at the Fe/Ge Interface}

\author{J.W. Freeland$^1$,  R.H. Kodama$^2$, M. Vedpathak$^2$, S.C. Erwin$^3$, D.J. Keavney$^1$, R. Winarski$^1$, P. Ryan$^1$, and R.A. Rosenberg$^1$,}
\affiliation{$^1$Advanced Photon Source, Argonne National Laboratory,
Argonne, IL.\ 60439}
\affiliation{$^2$Department of Physics,
University of Illinois at Chicago, Chicago, IL.\ 60607}
\affiliation{$^3$Center for Computational Materials
Science, Naval Research Laboratory, Washington, D.C. 20375}

\begin{abstract} We report direct experimental evidence showing induced magnetic moments on Ge at the interface in an Fe/Ge system. Details of the x-ray magnetic circular dichroism  and resonant magnetic scattering at the Ge L edge demonstrate the presence of spin-polarized {\it s} states at the Fermi level, as well as  {\it d} character moments at higher energy, which are both oriented antiparallel to the moment of the Fe layer. Use of the sum rules enables extraction of the L/S ratio, which is zero for the {\it s} part and $\sim0.5$ for the {\it d} component. These results are consistent with layer-resolved electronic structure calculations, which estimate the {\it s} and {\it d} components of the Ge moment are anti-parallel to the Fe {\it 3d} moment and have a magnitude of $\sim0.01\ \mu_B$.
 \end{abstract} 
%%75.70.Cn - % Interfacial magnetic properties % 75.30.Pd - Surface Magnetism % 
%%%79.60.Jv photoemission - Interfaces; heterostructures;nanostructures
\maketitle

Transport of spins across the ferromagnet/semiconductor interface and
their manipulation poses many fundamental questions but has wide
implications for spin-based electronics\cite{spintronics}. A key area is
understanding how spin polarized electrons or holes are injected from a
magnetic material into a semiconductor\cite{spininj1,spininj2}.
Semiconductor-based systems are intriguing since the electron's spin can
maintain coherence in a semiconductor while being transported over
length scales that are orders of magnitude larger than those achievable
in a metal\cite{spincoh}.   The Fe/GaAs system initially
demonstrated low efficiency of spin injection,\cite{fegaasspininj} which
was understood to result from the large conductivity mismatch between
the metal and semiconductor\cite{condtheory}. Recent results though have
demonstrated high spin injection efficiencies  by tunneling through the Schottky barrier formed at
the metal/semiconductor interface\cite{nrlapl}. However, to fully
understand these systems requires a detailed understanding of the
electronic and magnetic structure at this boundary, which will impact 
the ability for electron or holes to be transported into the
semiconductor.

Interest in Fe/semiconductor systems (e.g.,Fe/Ge, Fe/GaAs,\ldots) has been driven by the close
lattice match between the two materials enabling growth of single-crystal structures.  Early work demonstrated the ability to grow
expitaxial thin films\cite{fegaasprinz}, but significant intermixing at
the interface was observed as well\cite{fegexps,fegaasxps}. Further
studies demonstrated that at the thin film limit the Fe layer possessed
an in-plane uniaxial
anisotropy\cite{fegaasjonker,fegenorton,fegaasbland}, which has been
connected to thickness dependent interface strain\cite{fegaasscatt}.
Evidence of significant {\it 3d} charge transfer at the Fe/GaAs interface has
been observed and was connected to bonding between Fe and
As\cite{fegaasjwf}. With respect to the interface induced moment,
electronic structure calculations have predicted an induced moment and
modified density of states at the
boundary\cite{fegetheory1,fegetheory3,fegaastheory}. To date
though there has been no direct evidence for these induced interface
moments.

In this Letter we present direct evidence of spin-polarized Ge at the
interface with a magnetic transition metal. X-ray magnetic circular
dichroism at the Ge L edge in an Fe/Ge multilayer provided
unambiguous evidence for induced spin on Ge. Comparison with single
crystal results enable the identification of the spin polarization into
a {\it s} moment close to the Fermi level and a {\it d} component at
higher energy. The data is consistent with an antiparallel induced
moment on Ge with indications of a nonzero orbital moment for the {\it d}
component. Layer-resolved electronic structure calculations show the induced moment is localized around the interface with the {\it s} and {\it d} moments $\sim0.01\ \mu_B$ that are anti-parallel to the Fe {\it 3d} moment.

\begin{figure}[h!] 
\includegraphics[scale=.5]{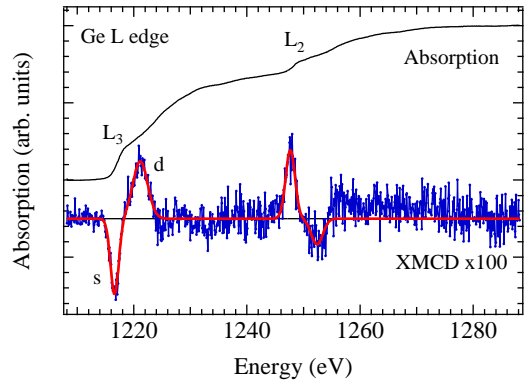}
\caption{Average absorption (I$^+$+I$^-$) and x-ray magnetic circular dichroism (XMCD = (I$^+$-I$^-$)) at the Ge L edge for the Fe/amorphous Ge multilayer. Nonzero XMCD gives clear indication of an induced magnetic moment on Ge. Structure in the XMCD is attributed to both spin-polarized {\it s } and {\it d} states, as well as a small magnetic signal in the extended signal, which might be due to magnetic extended absorption fine structure. The red line is a Gaussian fit to the data discussed in the text.}
\protect\label{gexmcdfe} 
\end{figure} 
Element-selective magnetic measurements were made using x-ray magnetic
circular dichroism (XMCD)\cite{xmcd}, which arises from the coupling of
the magnetic orientation to the x-ray polarization. These experiments
were performed at sector 4 of the Advanced Photon Source\cite{sect4rsi}.
Beamline 4-ID-C provides high-resolution
polarized x-rays in the intermediate x-ray range of 500 - 3000
eV. The x-rays are generated by a novel circularly polarized
undulator that provides left- and right-circular polarization switchable
at demand at a polarization $>$ 96$\%$. The samples were studied by 
measurement of total fluorescence yield (TFY) using a microchannel plate
detector. Fluorescence yield was used both because of its bulk sensitivity, as well
as the ability to measure in an applied magnetic field. The measurement
involves changing the magnetization direction  at each energy point of
the absorption curve to measure the absorption with photon helicity and
magnetization parallel ($I^{+}$) and antiparallel ($I^{-}$). The sum
($I^{+}$-$I^{-}$) provides chemical information while the XMCD
($I^{+}$-$I^{-}$) is magnetic in origin. Since the XMCD signal is very small, it was 
confirmed that the XMCD changed sign upon reversal of the photon helicity. In addition, x-ray resonant magnetic scattering (XRMS) was utilized to determine the spatial location of the induced spin\cite{xrms,jwfrc}.
\begin{figure}[h] 
\includegraphics[scale=.5]{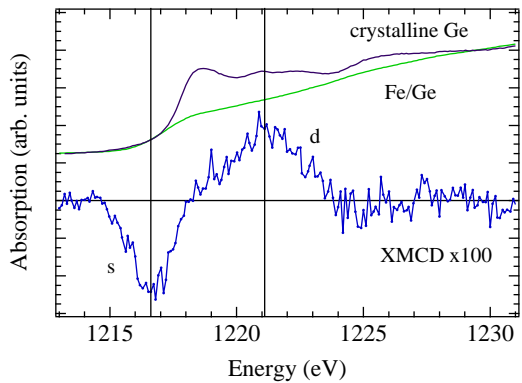}
\caption{ Comparison of Ge L$_{3}$ absorption for the Fe/Ge sample and a single-crystal
(100) standard. Note the distinct drop in the absorption for
the amorphous case, which is associated with the disorder of the
amorphous state. Lines indicate the alignment of XMCD features with features ascribed to {\it s} and {\it d} states in the absorption.} 
\protect\label{geabscmp} 
\end{figure}

To maximize the interface contribution to the absorption signal, a
multilayer sample was used. The Fe/Ge multilayers were prepared by
sputtering at room temperature with a nominal structure of Si(100)/Ge(100\
\AA)/[Fe(22\ \AA)/Ge(22\ \AA)]$_{20}$/Ge(100\ \AA) as confirmed
by x-ray reflectivity and TEM. It should
be noted here that Ge grown on Fe results in an amorphous
structure\cite{fegeml}.  With the Ge unit cell being 5.66\
\AA,  each monolayer (ML) will correspond to $\sim$2\ \AA\ of Ge from
which it can be estimated that the interface comprises about 10$\%$ of
the sample. A recent study though claims that this system
cannot be considered simply as Fe/Ge. A large percentage of intermixing
leading to coexisting Fe and Fe$_{x}$Ge$_{1-x}$
regions\cite{fegestruct}. From the details of this study we can estimate 
that the alloy component would comprise at most 25$\%$ of the sample.

Measurement of the
polarization-dependent absorption at the Ge L edge (see Figure
\ref{gexmcdfe}) provides direct access to the electronic and magnetic
order of Ge. Due to the dipole selection rules for x-ray absorption,
this excitation details the {\it d} and {\it s} contributions to the
spin polarized density of unoccupied states.   Presence of a nonzero
XMCD  provides direct evidence of the spin-polarized Ge, but  more
detailed analysis of the data  requires an understanding of the
electronic character of the unoccupied states
contributing to the XMCD. As will be shown below, the  XMCD can be considered as consisting of 3 fundamental components: a negative peak close to the Fermi level ascribed to spin-polarized {\it s} states, higher lying spin-polarized {\it d} states, and an extended component that may be attributed to magnetic extended absorption fine structure\cite{mexafs}. 
To assign the XMCD to particular electronic states first consider the results for bulk Ge, which will provide an initial framework. 

Calculations for amorphous Ge (a-Ge) show the density of unoccupied states are nearly
the same as crystalline (c-Ge) with only small modifications\cite{agedos}. The peaks
in the c-Ge case result from both from the {\it s} and {\it
d} density of unoccupied states as well as multiple scattering from the
highly ordered crystal\cite{gelabs2}. The edge for the c-Ge case was described by an initial peak due to {\it s} states followed by a double-peak feature due to {\it d} states. Results for the a-Ge system could be reproduced by the same density of states simply by varying the degree of coherent multiple scattering to simulate the disordered amorphous state\cite{gelabs2}.
A comparison of our results with the crystalline case is shown in Fig.\ \ref{geabscmp} and agrees well with previous results.
Using this result enables an inital assignment based on the bulk electronic structure. The first XMCD peak in  Fig.\ \ref{geabscmp} lies in the pre-edge region close to the Fermi level, which may have important potential impact on spin-dependent transport. Bulk electronic structure in this region is predominately {\it s} character. The second XMCD peak coincides with the doublet assigned to {\it d} states. This provides an initial assignment as shown in Fig.\ \ref{geabscmp}, which is consistent with the analysis of the experimental data and the layer resolved density of state calculations, both of which are discussed below. 

%To confirm this ... layer resolved...

%\begin{figure}[h] 
%    \includegraphics[scale=.5]{gelayerdos.eps} 
%    \caption{ Interface vs.\ bulk density of states which show the interface an bulk electronic structure just above the Fermi level are fairly similar. The curves are offset for clarity. } 
%    \protect\label{gedos} 
%\end{figure} 

With an assignment of features, the dipole selection rules enable the
determination of orientation between the Fe and Ge moments. The XMCD
selection rules for {\it p} $\rightarrow$ {\it d} vs.\ {\it p}
$\rightarrow$ {\it s}\cite{ssumrules1,ssumrules2} allow {\it s} and
{\it d} contributions to the spin moment, $<S_z>$, to be written as:
\begin{equation} <S_z>=<S_z^d>-2<S_z^s>{{P_s} \over {P_d}}+{7 \over
2}<T_z> \label{eq:szcomp} \end{equation} where ${{P_s} \over {P_d}}$ is
the probability ratio for {\it s} vs.\  {\it d} excitation and $<T_z>$
is the magnetic dipole term. In accord with recent results, it can be assumed that 
${{P_s} \over {P_d}} \sim 1$\cite{ssumrules2} and  $<T_z>\ \simeq\ 0$ \cite{sumrules3}. The important point is 
that the selection rules result in a minus sign between the {\it d}
vs {\it s} XMCD for the same magnetic moment direction. Since the initial L$_{3}$ XMCD
is negative for Fe (not shown), this is direct
evidence of the antiparallel orientation of the Ge {\it s}
moment and the {\it 3d} transition metal moment. The {\it d} component has a positive signal indicating that the {\it d} component of the induced Ge moment is antiparallel to that of Fe as well. These results are in agreement with the electronic structure calculations presented below. 

Sum rule analysis provides more detailed insight into the magnetic
structure, but there are difficulties in the application to the case of
Ge. Without a continuum excitation background subtraction and knowledge of the number
of {\it 4s} and {\it d}  holes in the conduction band, extraction of the
spin and orbital moments is not possible. This is further complicated by
the issue that not all of the electrons in the conduction band are spin
polarized. As well there is charge transfer at the interface, which will
mix the {\it 3d} electrons from Fe with the {\it 4d} electrons from Ge.
However, the ratio, $R_{m} = \Delta A_{L_{3}}/\Delta A_{L_{2}}$ , where
$\Delta A_{L}$ is the area of the respective XMCD peak, can
provide the ratio of spin to orbital moment. Due to the small signal to
noise in the spectrum, the peaks were fit with Gaussians, and the fit
data were used for the energy integration (see Fig.\ref{gexmcdfe}).
With small signal to noise, the result can be lead astray by noise
contributions.  Since the purpose is to determine general trends in the
spin-orbit ratio, the fit works very well and allows  the
two contributions to be separated assuming no overlap of the relative features.

From the area of the separate components, we can determine the orbit to
spin ratio defined as: \begin{equation} {{\left\langle {L_z}
\right\rangle } / {\left\langle {S_z} \right\rangle }}\sim {4 \over
3}{{R_{m} + 1} \over {R_{m} - 2}}. \end{equation} \noindent In this case
we have not included factors for the different character of the states.
The results are shown in Table\ \ref{gexmcdresults}.
The numbers for ${\left\langle {L_z} \right\rangle } / {\left\langle
{S_z} \right\rangle}$ provide additional confirmation of the character
assignment. The {\it s} component results in an $\left\langle {L_z}
\right\rangle \sim 0$, and the second peak in the XMCD results in a
nonzero $\left\langle {L_z} \right\rangle $ consistent with a {\it d}
component. It is worth noting that the orbital moment is a significant
fraction of the total moment and perhaps plays a role in the uniaxial
part of the anisotropy for Fe/Ge(100)\cite{fegenorton}.
 \begin{table}[h!]
  \caption{Results of XMCD data analysis.} 
  \begin{tabular}{|c|c|c|c|c|}  
  \hline State   & $\Delta A_{L_{3}}$   &  $\Delta A_{L_{2}}$  & $R_{m}$ & ${\left\langle {L_z} \right\rangle } / {\left\langle {S_z} \right\rangle }$  \\ \hline 
  {\it s}  & -0.0090 & 0.0092 & -.98 & 0.008 \\ 
  {\it d}  & 0.0122 & -.0043 &-2.9 & 0.5 \\ \hline 
  \end{tabular}
  \label{gexmcdresults} 
  \end{table}

To interpret these results theoretically, we turn to density-functional
theory calculations. Two sets of calculations were performed to
elucidate different aspects of induced Ge moments in Fe: (1) several
specific Fe/Ge interface models representing varying degrees of Fe-Ge
intermixing and (2) fixed-spin-moment calculations for bulk Ge, for the
purpose of estimating the ratio between induced {\it p}-moments (which are
the largest) to {\it s} and {\it d} moments (which are measured by the L edge XMCD).  The
calculations were performed within the generalized-gradient
approximation, using projector-augmented-wave potentials
\cite{kresse93,kresse96}. The plane-wave cutoff and k-point sampling were
sufficient to converge all quantities to the precision given.

%We used a 3x3x3 supercell of bcc Fe containing one Ge atom to model the
%isolated Ge substitutional. The impurity creates very little lattice
%distortion: the nearest-neighbor Ge-Fe distance is only 1.5$\%$
%larger than a normal Fe bond.  The spin moment of the Ge impurity
%is about 0.1 $\mu_B$ (5$\%$ of the bulk Fe moment), and is oriented
%antiparallel to the Fe magnetization. This impurity moment has
%roughly 90$\%$ p-character, a result to which we will return later.
\begin{figure}[h] 
\includegraphics[scale=.5]{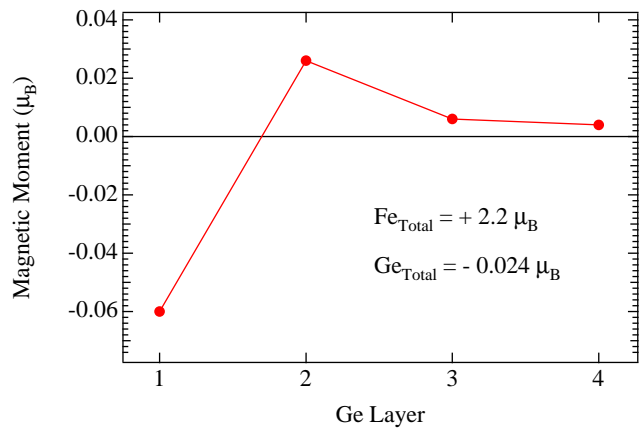}
\caption{ Layer-resolved calculated moments for an ideal Fe/Ge interface. The inset shows the average total moment for the structure, since XMCD in absorption is an average over all layers that contribute a magnetic signal.} 
\protect\label{fegemom}
\end{figure}
To study Fe/Ge interfaces, we used three models introduced previously
for Fe/GaAs interfaces\cite{fegaastheory}.  All are 1x1 interfaces between
crystalline Ge and Fe, with the interface boundary either atomically
abrupt or slightly intermixed (with 0.5 and 1.0 monolayers of Ge in the
Fe
host).  The isolated interfaces were modeled by supercells containing
seven layers each of Ge and Fe, with atomic coordinates completely
relaxed within two layers of the interface. The induced Ge moments for
the abrupt interface are shown in Fig.\ \ref{fegemom}; the results for the other two
models are qualitatively similar.  The interfacial Ge layer has a
moment of 0.06 $\mu_B$ antiparallel to the Fe magnetization.  Although
small, this is roughly half the value obtained for isolated Ge and
thus appears reasonable. Ge atoms one layer away from the interface are
less polarized (in the Fe direction), and further layers are
essentially nonmagnetic.   The Ge
moments near the interface are predominantly {\it p}-like.

However, the L-edge XMCD is only sensitive to the {\it s} and {\it d} components of the total Ge moment.  For
the calculations described above, these contributions were only $\sim10\%$
of the already small total Ge moment and hence are numerically
difficult to resolve.  To determine these ratios more systematically,
we performed fixed-spin-moment calculations for crystalline Ge,
varying the total moment, M, from zero to 1.5$ \mu_B$.  For each value of
M, we computed the{\it s}, {\it p}, and {\it d} contributions within an atomic sphere.
The ratios M$_s$/M$_p$ and M$_d$/M$_p$ are remarkably independent of the total
moment. They do, however, depend strongly on the sphere radius used:
for touching spheres M$_s$/M$_p$=0.3 and M$_d$/M$_p$=0.1, while for
volume-filling spheres M$_s$/M$_p$=0.1 and M$_d$/M$_p$=0.3.  Most importantly,
the {\it s}- and {\it d}-moments are always parallel to the {\it p}-moment.  These
results are completely consistent with the results obtained above for
the Fe/Ge interface. Hence, from these calculations, we conclude that the induced {\it s}- and {\it d}-moments on Ge atoms near the Fe/Ge interface are $\sim$ 0.01
 $\mu_B$, and are antiparallel to the Fe magnetization direction. However, since the XMCD averages over all layers, this measurement alone does not determine that the moment is localized at the Fe/Ge interface.
\begin{figure}[ht!] 
\includegraphics[scale=.5]{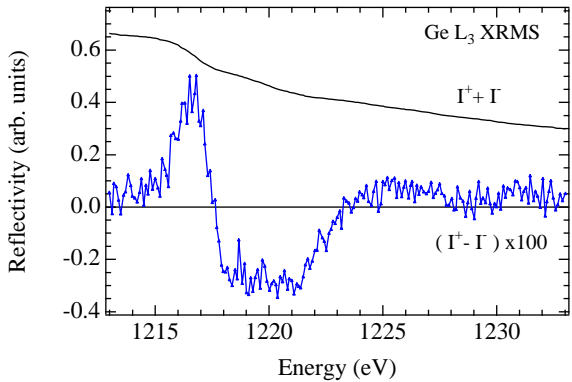}
\caption{ Average x-ray resonant scattering ($I^{+}$+$I^{-}$) and the charge-magnetic interference term ($I^{+}$-$I^{-}$) for an incident angle of 20$^o$ with respect to the plane of sample. Strong magnetic signature of the same magnitude provides indication that the induced magnetic moment coincides with the Fe/Ge interface.} \protect\label{gexrms}
\end{figure}

Indications of the localization of the Ge magnetic moment at the interface are provided by
measurements of the x-ray resonant magnetic scattering (XRMS).
Since XRMS probes changes in the chemical and magnetic density profile, 
it is very sensitive to the interface between two distinct elements. Figure\ 
\ref{gexrms} shows the average, as well as helicity-dependent signal
arising from the Ge moment. Since XRMS examines the density
profile, it is most sensitive to smooth interfaces with a distinct
change in the chemical profile. For the case of the Fe-Ge alloy, the 
interface will not be well defined and should not result in a strong 
signal. The results here show a strong magnetic component that is of 
the same magnitude as the absorption result. This is a good
indication of the moment localized near the Fe/Ge interface. Questions remain though concerning the presence of intermixed regions and how such a region would contribute to the magnetic signal. 
\begin{figure}[h!] 
\includegraphics[scale=.5]{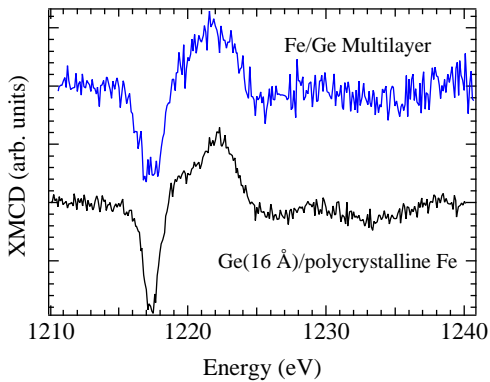}
\caption{ Comparison of XMCD from Fe/Ge multilayer and a 16\ \AA\ Ge film grown {\it in situ} on polycrystalline Fe. Within the noise, the two spectra are identical. Since the later shows minimal alloying of Fe, this is a good indication that the magnetic component resides at an Fe/Ge interface and is not due to an Fe-Ge alloy phase.} \protect\label{gecmp}
\end{figure}

 To confirm our result and rule out the possibility of this magnetic phase being ascribed to an alloy , an amphorous Ge thin film was  prepared {\it in situ} on a clean polycrystalline Fe surface. As mentioned above, a previous study of Fe/Ge/Fe structures provided evidence for the formation of an Fe-Ge alloy phase\cite{fegestruct}. The results of this paper concluded that the alloy formation occurred during the Fe on Ge growth, not for the Ge grown on Fe. Using a UHV surface science chamber, a sample with the following structure was prepared: Si/polycrystalline Fe(50\ \AA)/amorphous Ge(16\ \AA). X-ray photoelectron spectroscopy (XPS) showed the sample was free from contamination and there was minimal intermixing of Fe and Ge.  The measured XMCD spectra of the single Ge layer and  the Fe/Ge multilayer are  identical (see Fig.\ \ref{gecmp}), which proves that the moment is formed at the Fe/Ge interface and not due to an Fe$_x$Ge$_{1-x}$ phase. Future work will focus on these single-layer structures to extract more a more detailed understanding of the interface electronic and magnetic structure.

Use of the Advanced Photon Source was supported by the U.S. Department
of Energy, Office of Science, under Contract No.  W-31-109-Eng-38. Computations were performed at the DoD Major Shared Resource Centers at ASC. This work was supported in part by the Office of Naval Research.

\bibliographystyle{apsrev} 
\bibliography{GeXMCDBibfileetal}

\end{document}